\documentstyle[12pt]{article}
\begin{document}
\parskip 10pt plus 1pt
\title{Spacetime Dependent Lagrangians and Electrogravity Duality}
\author{
{\it Debashis Gangopadhyay $^{a}$,R.Bhattacharyya $^{b}$and  L.P.Singh $^{c}$}\\
{\it $^{a}$ S.N.Bose National Centre For Basic Sciences,}\\
{\it JD-Block,Sector-III,Salt Lake, Kolkata-700098,India}\\
{\it debashis@bose.res.in}\\
{\it $^{b}$ Dept.of Physics,Dinabandhu Andrews College,Kolkata-700084,India}\\
{\it $^{c}$ Dept.of Physics,Utkal University,Bhubaneswar-751004,India}
}
\date{}
\maketitle
\baselineskip=20pt

\begin{abstract}
We apply the spacetime dependent lagrangian formalism [1] to the action in 
general relativity. We obtain a Barriola-Vilenkin type monopole solution by exploiting the
electrogravity duality of the vacuum Einstein equations and using
a modified definition of empty space. An {\it upper bound} is obtained on the monopole
mass ${\tt M}$, 
${\tt M}\leq e^{(1-\alpha)/\alpha}/(1-\alpha)^{2}{\tt G}$ where $\alpha = 2k $
is the global monopole charge.



\end{abstract}

\newpage
{\bf 1.Introduction}

It has been shown that electromagnetic duality as well as weak-strong
duality in equations of motion can be obtained by 
introducing explicit spacetime dependence of the lagrangian [1]. Duality symmetry,
which hinges on nonperturbative aspects of a theory, is implemented through
certain general set of characteristics of a spacetime dependent function 
-representing the spacetime dependence of the    
lagrangian- at large values of the spacetime coordinates i.e. on the boundary. 
The solutions for the fields 
obtained under these conditions are topological in nature.
Electrogravity duality is a topological defect [2] generating process. Under electrogravity
duality transformations on the Einstein equations, vacuum solutions get 
mapped onto solutions with a global monopole.An example of a global monopole solution is 
the Barriola-Vilenkin (B-V)monopole [3].In this work we show that,
using the formalism of [1], one can get a B-V type
metric in asymptotic regions defined by $r\approx e^{1/\alpha}$
where the global monopole charge $\alpha$ is very close to zero.
We also additionally obtain an {\it upper} bound for the monopole mass:
${\tt M}\leq e^{(1-\alpha)/\alpha}/(1-\alpha)^{2}{\tt G}$ where ${\tt G}$ is 
Newton's gravitational constant. Interestingly one also finds an analogue of
the holographic principle operating in this example.

Vacuum Einstein equations are invaiant under  $G_{ik}\leftrightarrow R_{ik}$ 
where $G_{ik}= R_{ik} - (1/2)g_{ik}R$  is the Einstein tensor, $R_{ik}$ is Ricci 
tensor,$R$ is the Ricci scalar and $G$ the Einstein scalar. 
This signifies the existence of electrogravity duality symmetry.
Like the electromagnetic field, the gravitational field 
can be resolved into "electric" (due to charge) and "magnetic" (motion of charge) 
parts. For gravity the analogues are mass-energy and its motion respectively. 
Gravity entails two kinds of charges-- non-gravitational matter-energy ("active" part)
and gravitational field energy ("passive" part).Electrogravity duality means 
that vacuum equations are invariant under the interchange of the active and 
passive "electric" parts. Mathematically this entails interchange of Ricci and 
Einstein curvatures. Under such electrogravity duality transformations, vacuum solutions
get mapped onto topological solutions. 

In this context,there is  a definition of vacuum (empty space) [4]
which is less restrictive than $R_{ik}=0$ (i.e. vanishing of all Ricci components).
Vacuum, instead, is characterised by energy density 
(relative to a static observer) $\rho=0$, timelike convergence density
(relative to a static observer) $\rho_{t}=0$, null convergence density 
(relative to radial null geodesic) $\rho_{n}=0$ and the absence of any energy 
flux $P_{c}=0$. For a spherically symmetric metric $\rho_{n}=P_{c}=0$ imply
$R_{01}=0\enskip , \enskip R^{0}_{0}=R^{1}_{1}$ which in turn
implies that $g_{00}=f(r)=-g^{11}$; while $\rho=0$ means $R^{2}_{2}=0$ which 
integrates to give the Schwarzschild solution. So empty space is characterised
by
$$R^{2}_{2} =0=R^{0}_{1} \enskip ;  R^{0}_{0}=R^{1}_{1}\enskip\eqno(1a)$$
It has been shown [4] that if we replace $\rho$ by $\rho_{t}$ 
and use $\rho_{t}=\rho_{n}=P_{c}=0$, then this implies that 
$$G^{2}_{2} =0=G^{0}_{1}\enskip ; G^{0}_{0}=G^{1}_{1}\enskip \eqno(1b)$$ 
Then  replacing the Ricci tensor by the Einstein tensor leads 
to the solution of the metric tensor as that of the Barriola-Vilenkin monopole. 
$G$ is the electrogravity dual of $R$ . So the global 
monopole metric is electrogravity dual of the Schwarzschild metric. 

In the presence of non-gravitational sources we take a modified action 
shown in equation $(4)$ and field equations obtained from 
equation (2). These are motivated by  Ref.[1a].

{\bf 2.Global monopole and an upper bound for the global monopole mass}

We first recall our formalism [1].
Let the lagrangian $L'$  be a function of fields $\eta_{\rho}$, their derivatives 
$\eta_{\rho,\nu}$ {\it and the spacetime coordinates $x_{\nu}$}, i.e. 
$L'= L'(\eta_{\rho},\eta_{\rho,\nu}, x_{\nu})$.Variational principle yields 
$\int dV \left(\partial_{\eta}L'- \partial_{\mu}\partial_{\partial_{\mu}\eta}L'\right) = 0$.
Assuming a separation of variables : 
$L'(\eta_{\sigma},\eta_{\sigma,\nu},.. x_{\nu})=\Lambda(x_{\nu}) L(\eta_{\sigma},\eta_{\sigma,\nu})$, ($\Lambda(x_{\nu})$ 
is the $x_{\nu}$ dependent part and is a finite non-vanishing function) gives 
$$\int dV \left(\partial_{\eta}(\Lambda L)- \partial_{\mu}\partial_{\partial_{\mu}\eta}(\Lambda L)\right) 
= 0\eqno(2)$$
The usual action in gravity is (in units where velocity of light $c=1$)
$$S=  -(1/16\pi{\tt G})\int d^{4}x R \sqrt {-g} + (1/2)\int d^{4}x g^{ik} T_{ik} \sqrt {-g}\eqno(3)$$ 
$g$ is the determinant of the metric tensor,  ${\tt G}$ is Newton's 
gravitational constant and $T_{ik}$ the energy-momentum tensor. Consider the modified action 
$$S_{\Lambda}= -(1/16\pi{\tt G})\int d^{4}x \Lambda (x) R \sqrt {-g}\eqno(4)$$
In our formalism this will lead to equations of motion which when solved in the light of Dadhich's
approach yields a solution for the metric for a global monopole as well as $\Lambda$. 
The spacetime dependence (after the overall dependence on $R$) is expressed 
by $\Lambda$ (assumed to be a function of $r$ only; $r=\infty$ is a
boundary of the theory). $\Lambda$ is {\it not} dynamical and is a finite,
non-vanishing function at all $x_{\nu}$. It is like some external field and 
equations of motion for $\Lambda$ meaningless at the length scales (classical
gravity) under consideration. $\Lambda$ is finite at infinity. The
finite behaviour of $\Lambda$ on the boundary {\it encodes the electrogravity 
duality of the theory within the boundary} providing an analogue of  
't Hooft's holographic principle [5]. 

Using the action as defined in  $(4)$ the equations of motion that follow using $(2)$ are
$$\Lambda [R_{ik}-(1/2)g_{ik}R] -(\delta/\delta g^{ik})
[(\partial_{m} \Lambda)
(g^{pm}\delta\Gamma^{l}_{pl}-g^{pq}\delta\Gamma^{m}_{pq})]=0$$
where $\Gamma^{i}_{jk}$ are the affine connections. This can be recast into the form
$$\Lambda (R_{ik} - (1/2)g_{ik} R) - M_{ik} = 0$$
i.e.
$$\Lambda G_{ik} = M_{ik}\eqno(5a)$$
with
$$M_{ik}=(1/2)(g^{mn}\partial_{n}g_{ik}\partial_{m}\Lambda)
-\partial_{n}(g^{mn}g_{ik}\partial_{m}\Lambda)$$
$$-(g^{mn}\partial_{m}g_{kn}\partial_{i}\Lambda)
+(1/2)(g^{mn}\partial_{k}g_{mn}\partial_{i}\Lambda)
+\partial^{n}(g_{mi}g_{nk}\partial^{m}\Lambda)\eqno(5b)$$
$M_{ik}$ is like an effective stress tensor. 

We shall take the spherically symmetric metric
$$ds^{2} = B(r) dt^{2} - A(r) dr^{2} - r^{2}(d\theta^{2} 
+ sin^{2}\theta d\phi^{2})\eqno(6)$$
(i.e.$g_{00}=B(r)$ ; $g_{11}=- A(r)$).
Then 
$$G^{2}_{2}=M^{2}_{2}/\Lambda=({1\over Ar}){\partial_{1}\Lambda\over\Lambda} 
-({1\over A^{2}})(\partial_{1}A)({\partial_{1}\Lambda\over\Lambda})
+({1\over A}){\partial_{1}^{2}\Lambda\over\Lambda}\eqno(7a)$$
$$G^{0}_{0}=M^{0}_{0}/\Lambda=[({1\over A^{2}})\partial_{1} A 
-({1\over 2AB})(\partial_{1}B)]{\partial_{1}\Lambda\over\Lambda} 
-({1\over A}){\partial_{1}^{2}\Lambda\over\Lambda}\eqno(7b)$$
$$G^{1}_{1}=M^{1}_{1}/\Lambda=[-({1\over A^{2}})\partial_{1} A 
-({1\over 2AB})(\partial_{1}B)]{\partial_{1}\Lambda\over\Lambda} 
-({2\over Ar}){\partial_{1}\Lambda\over\Lambda} \eqno(7c)$$
$\Lambda$ is determined such that $(1b)$ is obeyed which translated in our
formalism means ($M_{ik}$ as defined in equation $(5b)$)
$M^{2}_{2}= 0\enskip ; \enskip M^{0}_{0}=M^{1}_{1}=-\alpha\Lambda/r^{2}$ i.e.
$$M^{2}_{2}=({1\over Ar})\partial_{1}\Lambda 
-({1\over A^{2}})(\partial_{1}A)(\partial_{1}\Lambda)
+({1\over A})\partial_{1}^{2}\Lambda=0\eqno(8a)$$
$$M^{0}_{0}=[({1\over A^{2}})\partial_{1} A 
-({1\over 2AB})(\partial_{1}B)]\partial_{1}\Lambda 
-({1\over A})\partial_{1}^{2}\Lambda=-({\alpha\Lambda\over r^{2}})\eqno(8b)$$
$$M^{1}_{1}=[-({1\over A^{2}})\partial_{1} A 
-({1\over 2AB})(\partial_{1}B)]\partial_{1}\Lambda 
-({2\over Ar})\partial_{1}\Lambda=-({\alpha\Lambda\over r^{2}}) \eqno(8c)$$

We shall now obtain solutions of equations $(8)$ under certain physically
plausible approximations. Specifically we show below (Case 2) that there
exists a region of large
$r$ given by $r\sim e^{1/\alpha}$, $\alpha\rightarrow 0$, where one has the 
Barriola-Vilenkin metric for  $\Lambda\not= constant$ (Case 2).

{\bf Case 1: $r=\infty$ ; $\alpha=0$ ; $\Lambda= constant$}

Then $M_{ik}=0$  for all $i, k$ and $(5a)$ reduces to the usual equation in 
general relativity. The solution is the usual Schwarzschild solution and 
global monopole solution is not possible.On the boundary ($r=\infty ; \Lambda=constant$)
the action is the usual one 
$$ S=-(1/16\pi {\tt G})\int d^{4}x R \sqrt -g  $$ 
This is in terms of the Ricci tensor whose solution gives the Schwarzschild solution.

{\bf Case 2: $r\not=\infty ; \alpha\not= 0 ; \Lambda\not= constant$}

Then $M_{ik}\not= 0$.Postulate $\Lambda = \Lambda(r)$ for $\alpha\not= 0$ .
We now illustrate that there exists a form for $\Lambda (r)$ which can lead
to the B-V monopole under certain assumptions. Consider 
$$\Lambda (r)={1\over (1-\alpha)}ln [r(1-\alpha) - 2{\tt GM}]\eqno(9)$$
In $(9)$, {\it $\alpha$ is never unity} so that $\Lambda$ is always
well defined. Also , {\it $\alpha$ is never zero}, because then we would have 
$\Lambda = constant$ (Case 1). Further, we shall take $\alpha$ to be small.
So $0<\alpha<1$.
 
The equation $(8a)$ implies $\partial_{1}\Lambda(r)= P_{1} A(r)/r$. 
Choose the constant of integration $P_{1}=1$.
For $\Lambda$ as in $(9)$ then implies
$$A(r)={1\over (1-\alpha-{2{\tt GM}\over r})}$$
Adding $(8b)$ and $(8c)$ and simplifying gives
$$ln B= ln [r(1-\alpha)- 2{\tt GM}] - 2 ln r$$ 
$$+({2\alpha\over (1-\alpha)})\int dr ({1\over r}) ln [r(1-\alpha)-2{\tt GM}]$$ 
For large $r$,
$$ln [r(1-\alpha)-2{\tt GM}]\approx ln(1-\alpha) - ln r - {2{\tt GM}\over r(1-\alpha)}$$
Carrying out the integrations and choosing the integration constants to be zero give:
$$ln B(r)= ln(1-\alpha)-{2{\tt GM}\over r(1-\alpha)} - ln r 
+{4\alpha {\tt GM}\over (1-\alpha)^{2} r} $$
$$ + {2\alpha ln(1-\alpha)\over (1-\alpha)} ln r +{\alpha\over (1-\alpha)}(ln r )^{2} $$
Now $ln r$ is never zero. The contribution from  the terms proportional 
to $ln r$ will be negligible in the region where
$$ln r= {(1-\alpha)\over\alpha}-2 ln (1-\alpha)\eqno(10a)$$
i.e.
$$ r= {e^{(1-\alpha)/\alpha}\over  (1-\alpha)^{2}}\eqno(10b)$$
Then $r\approx e^{1\over\alpha}$
for small $\alpha$ so that $\alpha^{2}$ 
and higher orders are negligible.{\it This is the region of large $r$}.
In this region it is easy to see that 
$$ln B(r)=ln (1-\alpha) - {(1-2\alpha)\over (1-\alpha)}[2{\tt GM}/r(1-\alpha)]$$
$$\approx ln(1-\alpha)+ln [1-2{\tt GM}/r(1-\alpha)]$$
$$\approx ln[(1-\alpha)(1-2{\tt GM}/r(1-\alpha)]
\approx ln[1-\alpha-2{\tt GM}/r]$$
so that $B(r)\approx  1-\alpha-2{\tt GM}/r$. Here we have assumed
that $\alpha{\tt G}$ is of second order of smallness compared to 
$\alpha$.{\it Therefore for $\alpha\not= 0$
but very small, $\Lambda$ as in (9), and asymptotic region defined by 
$r\approx e^{1/\alpha}$ we get the Barriola-Vilenkin metric for a global monopole:}
$$A(r)={1\over (1-\alpha-{2{\tt GM}\over r})}\eqno(11a)$$
$$B(r)\approx  1-\alpha-2{\tt GM}/r\eqno(11b)$$
Given the way we have defined the asymptotic 
region, the statement "${2\tt GM/r}$ is small for large $r$"
now reads "${2\tt GM/e^{{1/\alpha}}}$ is small for large r (i.e.
$r\sim e^{{1/\alpha}}$) with $\alpha\rightarrow 0$."

Further justification of the solutions $(11)$ is as follows.
Subtracting $(8c)$ from $(8b)$ and simplifying gives 
$\Lambda=- {constant\over 3 r^{3}}$ and we now show that 
in the asymptotic region as defined by us this is readily 
consistent with the solution $(9)$. Expanding the logarithm
in $(9)$ upto the third order in $r$ gives
$$\Lambda={1\over(1-\alpha)} ln r + {1\over (1-\alpha)} ln (1-\alpha)
-{2\tt GM\over (1-\alpha)r}
-({1\over 2})[{2\tt GM\over (1-\alpha)r}]^{2}$$
$$-({1\over 3})[{2\tt GM\over (1-\alpha)r}]^{3} - ......$$
In the asymptotic region , $r\sim e^{1/\alpha}$, if we want
only the third order term to survive we must have the remaining
terms upto order two vanish. Plugging in the values
of $r$ and $ln r$ from $(10a,b)$ into the above equation
implies that
$$\alpha(1-\alpha)^{3}(2{\tt GM})^{2} 
+ 2\alpha(1-\alpha)^{2}e^{(1-\alpha)/\alpha}(2{\tt GM})$$
$$+2\alpha ln(1-\alpha)e^{2(1-\alpha)/\alpha}-2(1-\alpha)e^{2(1-\alpha)/\alpha}=0\eqno(12)$$ 
This is a quadratic in $(2{\tt GM})$.Remembering that both
$\tt G$ and $\tt M$ are non negative, and keeping terms upto
first order in $\alpha$ a solution is 
$$2{\tt GM}\approx e^{(1-\alpha)/\alpha}[(3/2)\sqrt {2\alpha} - 1]\eqno(13)$$
Certain points are to be noted. ${\tt G}$ is a universal positive
constant and ${\tt M}$ is also positive. This means that 
$\alpha\geq 2/9\approx 0.22222..$.Then if we take ,for example,
$\alpha\sim 0.22223$,we have 
${2{\tt GM}\over e^{(1-\alpha)/\alpha}}\sim 0.000017$.
The aim of this exercise is to
show that as per the definition of our asymptotic region 
${2{\tt GM}\over e^{(1-\alpha)/\alpha}}$ is indeed small even for a finite 
$\alpha$. Therefore in the asymptotic region the solution $(9)$ is 
consistent with the expression $\Lambda=-{constant\over 3r^{3}}$ where the 
the value of the constant is ${2\tt GM\over (1-\alpha)^{3}}$.
We have thus demonstrated unambiguously that the Barriola-Vilenkin type
metric can be obtained in our formalism under certain approximations.

Usually ${\tt GM}/r$ is taken to be a small quantity in general relativity 
for large values of $r$. So ${\tt GM} < \infty$. This scenario is for the 
usual Schwarzschild metric.For the electrogravity dual theory 
(in our spacetime dependent
lagrangian formalism) the asymptotic region is defined by 
$r= {e^{(1-\alpha)/\alpha}\over (1-\alpha)^{2}}$ with
$\alpha\rightarrow 0$. Then ${\tt GM}\leq {e^{(1-\alpha)/\alpha}\over (1-\alpha)^{2}}$ means 
$${\tt M}\leq {e^{(1-\alpha)/\alpha}\over (1-\alpha)^{2}{\tt G}}\eqno(14)$$.
This is an {\it upper bound} on the monopole mass and can be made very small 
by choosing 
$\alpha$ large. This is consistent with that of Harari and Lousto [3b], while
avoiding the physically undesirable feature of the monopole
mass becoming negative.

The expression $(14)$ reminds us of the Bogomolny bound 
${\tt M_{monopole}}\geq vg$
for usual monopoles where $v$ is the vacuum expectation value of the Higgs field and
$g$ the monopole charge, except that the Bogomolny bound is a {\it lower bound}
on the monopole mass.Carrying the analogy further, $(1/{\tt G})$ 
is like the vacuum expectation value of some field hitherto undiscovered while $\alpha=2k$ 
is defined as the  global monopole charge. 

The functional forms for $\Lambda$ are different in the two cases
and one cannot go from Case 1 to Case 2  ( or {\it vice-versa}) by 
varying $r$. The same $\Lambda$ cannot encompass both regions
by a variation of $r$.This is expected because Case 1 corresponds
to usual space whereas Case 2 is a consequence of electrogravity 
duality which is a topological defect generating process.

{\bf 3. The electrogravity dual action}

Consider now the respective actions.
On the boundary ($r=\infty$ ; $\alpha = 0$; $\Lambda=constant$) it is  the 
usual one :
$$S= -(1/16\pi{\tt G})\int d^{4}x  R \sqrt {-g}$$
This is  in terms of the Ricci tensor whose solution gives the
usual Schwarzschild solution.

Within the boundary, ($r\not=\infty$ ;$\alpha\not= 0$ ; $\Lambda\not=constant$)
{\it the action should be in terms of $M$} i.e. 
$$S_{\Lambda}^{dual} = -(1/16\pi{\tt G})\int d^{4}x M \sqrt {-g}$$
$$= -(1/16\pi{\tt G}) \int d^{4}x \Lambda G \sqrt {-g}
= -(1/16\pi{\tt G}) \int d^{4}x \Lambda (-R) \sqrt {-g}$$
$$= -(1/16\pi{\tt G^{dual}}) \int d^{4}x \Lambda R \sqrt {-g}\eqno(15)$$
{\it Therefore, the action in the dual theory ($\Lambda\not=constant$)
in our formalism must be defined with Newton's constant ${\tt G}$ replaced by
${\tt G^{dual}} = -{\tt G}$ . Then $(4)$ and
$(15)$ are equivalent definitions for the action and will lead to same field 
equations $(5a)$}. 

Apparently the Ricci tensor for the theory described by the metric $(11a,b)$ 
is $R=-M/\Lambda = 2\alpha/r^{2}$. 
However, taking into account the discussion following eq. $(15)$, the action 
should be defined with ${\tt G'=-G}$. With this definition the Ricci tensor for 
$(11a,b)$ is $R= - 2\alpha/r^{2}$. Alternatively, electrogravity duality 
interchanges $G$ and $R$, so the object to look for is $G$ i.e.
$G=-R=-2\alpha/r^{2}$. Thus whichever way one looks, things turn out to be consistent.

The metric also has a spacetime singularity at $r=0$
and represents a global monopole metric obtained as the dual of the Schwarzschild metric. 
On the boundary (Case 1),$\alpha = 0$, $\Lambda=constant$; one has the usual 
Schwarzschild metric. 
Within the boundary (Case 2) in the asymptotic region defined by 
$r\approx e^{1/\alpha}$,
$\alpha\not= 0, \Lambda\not = constant$; we have the electrogravity dual solution 
{\it viz.} the Barriola-Vilenkin metric. 
This again reminds us of the 't Hooft's holographic principle [5].

{\bf 4.Conclusion}

Thus the spacetime dependent lagrangian formalism can accommodate electrogravity
duality in all its aspects.
This is illustrated by obtaining the Barriola-Vilenkin monopole metric 
from the equations of motion under certain approximations.
No scalar field is necessary.$\alpha$ may as well originate
from other types of non-gravitational matter-energy 
sources hitherto unknown. A scalar field is merely one such realisation.
The solution follows from electrogravity duality 
which can be precisely formulated in our formalism 
by stating that {\it the solutions for  $\Lambda (r)\not= constant$ 
are the electrogravity dual of solutions for $\Lambda (r) = constant$.} 
An analogue of the holographic principle is
again illustrated. 
The global monopole mass is predicted to satisfy an {\it upper bound}.
Another point need to be stressed, {\it viz.} the empty space definition $(1)$. 
This definition is  naturally adaptable to our formalism which has 
already been shown to be successful in a full quantum theory exhibiting 
weak-strong duality [1b]. There seems to exist some deeper significance 
in this definition {\it vis-a-vis} quantum field theory.
This aspect requires further investigation.

\end{document}